\documentclass[12pt]{article}

\usepackage[margin=1in]{geometry}
\usepackage{graphicx}
\usepackage{booktabs}
\usepackage{amsmath}
\usepackage[numbers]{natbib}
\usepackage{xurl}
\usepackage[hidelinks]{hyperref}
\usepackage{subcaption}
\usepackage{caption}

\begin{document}

\title{From Agent Identity to Agent Economy: Measuring the Operational Readiness of ERC-8004 AI Agents}

\author{
\begin{tabular}{cc}
\begin{minipage}{0.45\textwidth}
\centering
Rischan Mafrur\href{https://orcid.org/0000-0003-4424-3736}{\,\textsuperscript{ORCID}}\\[0.4em]
{\large Western Sydney University}\\
{\scriptsize \texttt{r.mafrur@westernsydney.edu.au}}
\end{minipage}
&
\begin{minipage}{0.45\textwidth}
\centering
Priagung Khusumanegara\\[0.4em]
{\large Telkomsel}\\
{\scriptsize
\href{mailto:priagung_khusumanegara@telkomsel.co.id}
{\texttt{\nolinkurl{priagung_khusumanegara@telkomsel.co.id}}}
}
\end{minipage}
\end{tabular}
}

%\date{June 2026}

\maketitle

\begin{abstract}

This paper examines whether blockchain-registered AI agents demonstrate operational readiness beyond identity registration. Using a dataset of ERC-8004 agents on Ethereum, we construct an agent-level feature table covering identity status, metadata, service declarations, reputation feedback, transfers, and cross-chain registration. We develop an operational readiness framework based on observable evidence layers and complement it with network analysis of owner-agent, feedback-client, wallet-transfer, and combined evidence relationships. The results show that early ERC-8004 adoption is registration-heavy but operationally shallow. While the identity layer is visible at scale, metadata availability, service exposure, reputation formation, and cross-chain evidence remain limited. Ownership and feedback activity are also highly concentrated, suggesting that early participation is shaped by a small number of high-activity wallets and clients. The network analysis further shows that richer operational evidence clusters around a small subset of agents rather than being broadly distributed across the ecosystem. The findings suggest that ERC-8004 provides an important identity layer for decentralized AI agents, but the transition from agent identity to agent economy remains incomplete.

\end{abstract}


\section{Introduction}

Autonomous AI agents are increasingly discussed as future economic actors that can search for information, coordinate with other agents, execute transactions, and provide services with limited human intervention \citep{Tomasev2025,Xu2026}. In decentralized environments, this vision depends not only on advances in artificial intelligence, but also on infrastructure that allows agents to be identified, discovered, paid, evaluated, and trusted. Blockchain platforms such as Ethereum provide a natural setting for this infrastructure because they support machine-readable identity, programmable accounts, transparent transaction records, and public trust mechanisms \citep{Xu2026, Cao2022,AlJasem2025}. Within this setting, ERC-8004 \citep{DeRossi2025} has emerged as a proposed Ethereum standard for trustless AI agent infrastructure. It assigns each agent a unique on-chain identity, represented through an ERC-721 style token, and links this identity to metadata, and reputation. In principle, such infrastructure could support an agent economy in which autonomous agents offer services, transact with one another, build reputation, and participate in decentralized coordination \citep{Liu2026}.

Yet the existence of a technical standard does not by itself demonstrate that an agent economy is operating in practice. A registered agent may simply be an on-chain identity with little or no metadata, no declared service endpoint, no reputation history, and no evidence of interaction. This distinction is important because prior work on AI and blockchain warns that tokenized AI projects can create an illusion of decentralization when claims of autonomy, intelligence, or trust are not supported by verifiable usage, transparent execution, or effective governance \citep{Mafrur2025,Romandini2025,Yu2026}. In this context, the key empirical question is not whether ERC-8004 agents exist, but whether they show observable signs of operational readiness. Are registered agents discoverable? Do they expose services? Do they receive reputation feedback? Are ownership, feedback, and interaction patterns broadly distributed, or concentrated around a small number of wallets and clients?

This paper addresses these questions by examining the transition from agent identity to agent economy. Using a dataset of 10,000 ERC-8004 agents on Ethereum, we analyze covering identity status, metadata, service declarations, reputation feedback, transfer activity, and cross-chain registration. We then develop an operational readiness framework based on observable evidence layers. In this framework, an agent progresses from basic identity registration toward richer forms of readiness when it provides metadata, declares services, accumulates feedback, appears in cross-chain records, or participates in observable transfer activity. This approach allows us to distinguish between formal registration and more substantive evidence of ecosystem participation.

Our empirical analysis shows that early ERC-8004 adoption is registration-heavy but operationally shallow. Although the dataset contains 10,000 registered agents, only 67 agents expose service records, only 628 agents receive reputation feedback, and only 19 agents combine metadata, services, feedback, and cross-chain registration. Ownership and reputation activity are also highly concentrated: the top 10 owner wallets hold 51.40\% of agents, while the largest feedback client contributes 65.82\% of all feedback records. These findings suggest that ERC-8004 currently functions more strongly as an identity-registration layer than as a mature agent economy with broad service discovery, distributed reputation formation, and multi-party interaction.

To complement the readiness indicators, we also conduct a network analysis of the ERC-8004 ecosystem. We construct owner-agent, feedback-client-agent, wallet-transfer, and combined evidence networks to examine how agents are connected to wallets, feedback providers, service domains, and cross-chain records. The network results reinforce the main descriptive findings. Richer operational evidence is clustered around a small subset of agents, while most registered agents remain disconnected from the higher layers of service, feedback, and cross-chain activity. This network perspective is important because an agent economy should not only contain many registered identities, but should also exhibit distributed relationships among agents, users, service endpoints, and trust mechanisms.

This paper makes three contributions. First, it provides one of the first empirical assessments of ERC-8004 as a blockchain-based agent identity and trust standard. Second, it proposes a layered operational readiness framework for decentralized AI agents, linking identity, metadata, services, feedback, reputation, and cross-chain registration. Third, it introduces a network-based perspective on agent ecosystem maturity by visualizing how agents connect to owners, feedback clients, service domains, and other infrastructure signals. Together, these contributions provide a practical way to evaluate whether blockchain-registered AI agents are progressing beyond identity creation toward an operational agent economy.

\section{Literature Review}

\subsection{Autonomous AI Agents and Blockchain-Based Agent Infrastructure}

Recent advances in large language models and multi-agent systems have renewed interest in autonomous software agents that can interpret goals, reason over context, use tools, and act with limited human intervention \citep{FanMin2026,FanYang2025}. In conventional digital environments, these capabilities are often evaluated through task completion, planning accuracy, or tool-use performance. In Web3 environments, however, the requirements are more demanding. Agents must interpret on-chain state, interact with smart contracts, manage wallet-related operations, and avoid irreversible transaction errors \citep{FanMin2026,Romandini2025, Karim2025}. As a result, blockchain-based agents cannot be evaluated only by whether they generate plausible instructions or recommendations. They must also be assessed by whether they can operate safely, reliably, and transparently in transaction-based environments.

The existing literature illustrates both the promise and the limits of this transition. Language-model agents can translate natural language instructions into structured on-chain operations, but reliable execution depends on intent extraction, instruction decomposition, state grounding, calibration, and error handling \citep{FanMin2026}. Evidence from autonomous trading and financial-market agents also shows that general language-model capability does not automatically translate into reliable domain-specific decision-making \citep{FanYang2025}. These findings suggest that autonomy is not a binary property. An agent may be technically capable of generating actions, but still lack the surrounding infrastructure needed for safe and trustworthy participation in blockchain markets.

Security and privacy research reinforces this point. Blockchain AI agents can be understood along a spectrum from conversational systems, to instruction-following systems, to goal-directed systems that execute actions or manage objectives over time \citep{Romandini2025, Chaffer2024, Ranjan2025}. As agents move along this spectrum, risks increase. Goal-directed agents may introduce risks related to erroneous behavior, prompt injection, context manipulation, privacy leakage, wallet exposure, market instability, and loss of human oversight \citep{Romandini2025}. These risks make trust infrastructure central to any agent economy. Before agents can transact or coordinate at scale, other actors must be able to identify them, discover their services, assess their past behavior, and verify whether their outputs or claims are reliable.

Identity is therefore a foundational requirement for autonomous economic interaction. Humans rely on legal personhood, institutional records, and social identity to open accounts, enter contracts, and build reputation. Autonomous agents do not naturally possess these mechanisms. Instead, machine actors require cryptographic identity and verifiable credentials that can be interpreted by other machines \citep{Xu2026,Tomasev2025}. Ethereum's token standards provide part of this foundation. The non-fungible token model allows unique digital entities to be represented as distinct on-chain objects, making it suitable for persistent agent identities \citep{Entriken2018,Dafflon2023}. ERC-8004 extends this logic by proposing a registry-based architecture for trustless AI agent infrastructure \citep{DeRossi2025,Liu2026}.

ERC-8004 provides a framework through which agents can be represented, discovered, evaluated, and validated using Ethereum-based records. The standard is organized around three linked components: identity, feedback, and reputation. The Identity Registry provides each agent with a unique on-chain identity, represented through an ERC-721 style token. The Reputation Registry allows counterparties to submit feedback about agents. The significance of ERC-8004 lies in this connection between identity and trust. However, the existence of a standard does not mean that registered agents are operationally ready. A functioning agent economy would require agents to have accessible metadata, declared services, meaningful feedback and interaction patterns that extend beyond isolated minting. This distinction between formal registration and operational use motivates the empirical focus of this paper.

\subsection{From Agent Identity to Agent Economy: Feedback, Reputation, and Operational Readiness}

The idea of an agent economy has received increasing attention in recent work on decentralized AI and autonomous commerce. In this view, virtual agent economies are environments in which AI agents transact, negotiate and coordinate at speeds and scales beyond direct human oversight \citep{Tomasev2025}. Such systems require careful design around market permeability, reputation, resource allocation, credit assignment, and governance \citep{Tomasev2025}. Related work also argues that agents require a layered blockchain foundation that combines physical infrastructure, identity and agency, economic settlement, and collective governance \citep{Xu2026}.

Empirical studies of blockchain-based AI agents suggest that the ecosystem remains early, fragmented, and unevenly developed. Existing evidence shows that autonomous AI agent projects in decentralized finance already span trading, analytics, infrastructure, community engagement, and entertainment, but many of these projects remain heterogeneous and experimental rather than operationally mature \citep{Ante2026}. Studies of DeFi investment agents similarly show that tokenized agent markets can attract substantial valuations while providing limited evidence of verifiable autonomous execution, sustained performance, or stakeholder alignment \citep{Yu2026}. Related work on AI-based crypto projects further cautions that decentralization claims may become an illusion when core computation, governance remains off-chain and weakly verifiable \citep{Mafrur2025}. Taken together, this literature suggests that agent economies should not be inferred from branding, tokenization, or identity creation alone.

Operational readiness therefore provides a useful lens for distinguishing formal registration from actual ecosystem maturity. In blockchain settings, readiness implies active participation, observable use, and meaningful interaction rather than the mere existence of an on-chain record \citep{Yu2026}. In the AI agent context, this means that readiness should be understood as a progression from identity infrastructure toward economic functionality. An agent may first appear as a registered identity, but it becomes more operationally meaningful only when it provides accessible metadata, declares services, receives reputation feedback, participates in cross-chain or transfer activity, and generates evidence that other actors can inspect or evaluate. This paper builds on this logic by defining observable readiness layers and examining whether ERC-8004 agents move beyond identity registration toward richer forms of ecosystem participation.

A mature agent economy requires observable participation across multiple layers. Agents must be discoverable, able to expose services, accumulate reputation, and interact with other agents or users in ways that can be audited. Reputation is particularly important because it connects identity with historical performance. For autonomous agents, these mechanisms must be replaced or supplemented by machine-readable records that can be verified and reused across systems \citep{Tomasev2025,Xu2026}. On-chain reputation offers one possible approach because feedback can be recorded publicly, associated with a persistent agent identity, and queried by other participants.

At the same time, reputation systems must be interpreted carefully. A feedback record is useful only if it reflects meaningful interaction, credible evaluation, and sufficient independence between the evaluator and the agent. Existing work argues that agent markets require robust reputation mechanisms to support trust, specialization, and collaboration, and that historical performance can become a form of reputation capital or trust signal for autonomous agents \citep{Tomasev2025,Xu2026}. Related security research also emphasizes the need for accountability and auditability because blockchain actions can be irreversible and financially consequential \citep{Romandini2025}. These studies suggest that reputation should not be measured only by whether feedback exists. It should also be assessed by who provides feedback, how concentrated feedback activity is, what dimensions are being evaluated, and whether reputation is distributed across a wider ecosystem.

The literature also highlights broader risks that may affect the pace and shape of agent infrastructure adoption. Autonomous and decentralized agents may introduce security vulnerabilities, including prompt injection, key leakage, malicious tool use, and unintended transaction execution \citep{Romandini2025}. They may also create privacy risks when agents process sensitive user instructions, wallet data, or off-chain context \citep{Tomasev2025,Romandini2025}. In financial settings, autonomous agents could amplify market volatility if many agents react to similar signals, execute similar strategies, or interact through automated markets \citep{Ante2026}. These risks suggest that sparse adoption of agent infrastructure may reflect not only technical immaturity, but also caution around trust, governance, and safety.

The existing literature therefore leaves an important empirical gap. Prior work has developed conceptual models of agent economies, evaluated individual autonomous agents, and discussed the risks of blockchain-based AI agents. However, there is still limited evidence on whether the infrastructure proposed for agent economies is being used in practice. In particular, it remains unclear whether ERC-8004 agents are merely being registered as identities, or whether they are developing the metadata, services, reputation records, and network relationships needed for operational readiness. This paper addresses that gap by examining observable ERC-8004 activity on Ethereum mainnet. Instead of asking only whether agents exist, we assess whether their on-chain and metadata records show evidence of progression from agent identity toward an operational agent economy.

\section{Conceptual Framework}

This paper conceptualizes ERC-8004 agents through a layered operational readiness framework. The framework starts from a simple premise: an agent economy cannot be inferred from identity registration alone. A blockchain-registered agent may exist as an on-chain object, but this does not necessarily mean that it is discoverable, service-capable, trusted or economically active. Operational readiness therefore refers to the extent to which an agent progresses from basic identity registration toward observable forms of ecosystem participation. Building on prior work on agent economies, decentralized identity, reputation, and blockchain-based agent infrastructure \citep{Xu2026,Tomasev2025,Romandini2025}, we define readiness as the movement from agent identity toward usable and verifiable agent activity.

The framework contains several connected layers. At the foundation is identity readiness, where an agent is registered on-chain and assigned a persistent identity through an ERC-721 style token. This layer makes the agent visible, but it is only a minimal condition for participation. The next layer is metadata readiness, where the agent provides structured information such as a name, description, status, image, capability claims, or interaction-related attributes. Metadata is important because it links a cryptographic identity to a usable description of what the agent is and what it claims to do. The third layer is service readiness, where the agent declares endpoints, interfaces, or service capabilities that allow other actors to interact with it. This may include web endpoints, documentation links, communication protocols, or agent-native service formats such as MCP, A2A, OASF, and related endpoint structures. The fourth layer is reputation readiness, where feedback records connect the agent's identity and service claims to evidence of past interaction. Reputation is therefore a key bridge between registration and trust formation.

To operationalize this framework, identity registration is treated as the baseline condition because all agents in the sample are registered agents by construction. We then measure whether each agent shows additional observable evidence beyond this baseline. Specifically, we construct five Boolean indicators: \texttt{has\_metadata}, which captures whether an agent has observable metadata evidence; \texttt{has\_service}, which captures whether an agent declares at least one service record; \texttt{has\_feedback}, which captures whether an agent has received at least one feedback record; \texttt{has\_crosschain}, which captures whether an agent appears in cross-chain registration records; and \texttt{has\_transfer}, which captures whether an agent has observed transfer activity. These indicators correspond to observable forms of ecosystem participation, but they should be interpreted carefully. In particular, transfer activity indicates ownership movement or custody changes, not direct evidence of autonomous behavior.

We combine these indicators into an observable evidence score. The score is calculated as the sum of the five Boolean indicators, excluding identity registration itself. An agent with a score of zero is registered but has no additional observable evidence layer. An agent with a score of one has one additional evidence layer, such as metadata or feedback. Higher scores indicate that the agent is connected to multiple forms of observable ecosystem participation. This scoring approach provides a simple and transparent way to distinguish registration-only agents from agents with richer operational signals.

\section{Data and Methodology}

We analyze a compiled dataset of ERC-8004 agents introduced by \citet{Liu2026, Liu2026Data}. The sample consists of agent IDs 0--9999, covering the first 10,000 agents registered on Ethereum mainnet. The on-chain data were collected from block 24,339,925 on January 29, 2026 to block 24,839,925 on April 9, 2026. The dataset is provided as CSV tables: \texttt{agents\_core}, \texttt{agent\_metadata}, \texttt{agent\_services}, \texttt{agent\_reputation\_summary}, \texttt{agent\_feedback\_records}, \texttt{mint\_economics}, \texttt{transfer\_history}, and \texttt{crosschain\_registrations} \citep{Liu2026, Liu2026Data}. These tables allow us to reconstruct agent-level records covering identity, metadata, services, reputation, minting, transfer, and cross-chain evidence.

We perform data cleaning and integration as follows. All Ethereum addresses are lowercased to ensure consistent matching across tables. Boolean fields, such as \texttt{active\_flag}, are converted into Boolean values. For each agent, we merge information from the raw tables into a single feature record. We extract agent-level counts and indicators, including the number of service records, the number of unique service types, the number of feedback records, the number of unique feedback clients, and feedback values in both raw and decimal formats. We also extract ownership and transfer-related variables, including owner wallet, transfer count, and recipient diversity. Ownership concentration metrics, including the Herfindahl-Hirschman Index (HHI) and Gini coefficient, are computed from the distribution of agent counts across owner wallets.

Building on the integrated feature table, we construct agent-level readiness indicators. Identity registration is treated as the baseline condition because all agents in the sample are registered agents by construction. We then define additional observable evidence indicators: \texttt{has\_metadata}, which captures whether an agent has observable metadata evidence; \texttt{has\_service}, which captures whether an agent declares at least one service record; \texttt{has\_feedback}, which captures whether an agent has received at least one feedback record; \texttt{has\_crosschain}, which captures whether an agent appears in cross-chain registration records; and \texttt{has\_transfer}, which captures whether an agent has observed transfer activity. These indicators allow us to distinguish agents that remain at the level of basic identity registration from those that show additional signs of ecosystem participation.

We combine these indicators into an observable evidence score. The score is calculated as the sum of the five Boolean indicators, excluding identity registration itself. An agent with a score of zero is registered but has no additional observable evidence layer. Higher values indicate that the agent is connected to multiple forms of observable activity, such as metadata, services, feedback, cross-chain registration, or transfer activity. This score provides a transparent way to measure whether ERC-8004 agents are progressing from identity registration toward operational readiness.

In addition to the descriptive readiness analysis, we estimate an exploratory logistic regression to examine which observable agent characteristics are associated with reputation formation. The dependent variable is \texttt{has\_feedback}, which equals one if an agent has received at least one feedback record and zero otherwise. This outcome is used because feedback is one of the clearest observable signals that an agent has entered the reputation layer of ERC-8004. The model is specified as follows:

\begin{equation}
\Pr(\texttt{has\_feedback}_i = 1) =
\Lambda(\alpha + \beta X_i + \gamma Z_i),
\end{equation}

where \(\Lambda(\cdot)\) denotes the logistic function, \(X_i\) represents readiness-related variables for agent \(i\), and \(Z_i\) represents controls related to minting and ownership. The readiness variables include metadata indicators, service indicators, cross-chain registration, transfer activity, active metadata status, and lifecycle metadata linkage. The minting and ownership controls include log-transformed mint cost, gas used, mint batch size, and owner-group categories.

The regression is included as a supplementary association analysis, the purpose is to identify whether agents with richer observable infrastructure are more likely to receive feedback. This is important because the descriptive analysis shows how many agents reach each readiness layer, while the regression examines which observed features are associated with entry into the reputation layer. We use log-transformed variables for skewed count and cost measures, including service records, unique service types, transfer counts, minting cost, gas used, and mint batch size. A balanced logistic regression is used to account for the imbalance between agents with and without feedback. The coefficients are interpreted as associations conditional on the other variables in the model and should not be interpreted causally, since feedback may be shaped by off-chain review processes, dominant clients, platform-specific mechanisms, or early-stage ecosystem sparsity.

Finally, we complement the readiness indicators and regression analysis with network analysis. The network analysis examines whether ERC-8004 activity is broadly distributed across many participants or concentrated around a small number of wallets, clients, service domains, and cross-chain records. This is important because a mature agent economy should not only contain many registered identities, but should also exhibit distributed relationships across identity, service, reputation, and interaction layers.

% ============================================================
% Figure: Operational readiness funnel
% Draft label: fig:readiness_funnel
% File exists: 01_operational_readiness_funnel.pdf
% ============================================================

\begin{figure}[htbp]
    \centering
    \includegraphics[width=0.9\textwidth]{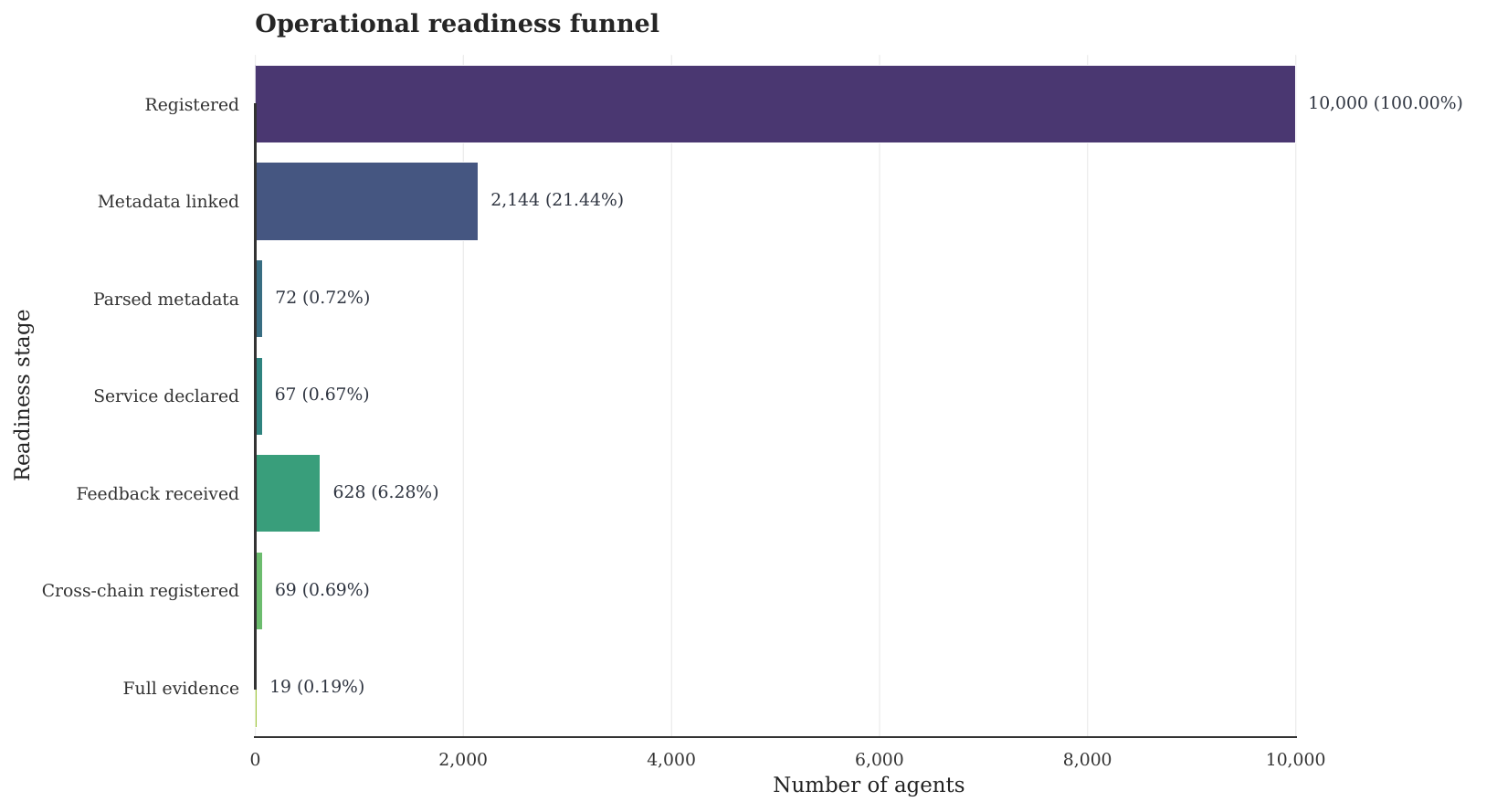}
    \caption{Operational readiness funnel for ERC-8004 agents. The figure shows how the sample narrows from identity registration to metadata, services, feedback, and cross-chain registration. The final category, full evidence, refers to agents that simultaneously have observable metadata, at least one service record, at least one feedback record, and at least one cross-chain registration record.}
    \label{fig:readiness_funnel}
\end{figure}

\section{Results}

\subsection{Descriptive Analysis}

The descriptive results reveal a substantial gap between the conceptual promise of ERC-8004 and its current empirical implementation. Although identity registration is clearly observable, with 10,000 agents created and discoverable on-chain, the additional dimensions required for operational readiness remain weakly developed. This finding is consistent with \cite{Mafrur2025}, which cautions that blockchain-based AI projects may appear decentralised at the architectural or narrative level while lacking substantive ecosystem development in practice. In the case of ERC-8004, the evidence suggests that registration alone does not translate into meaningful participation, service provision, or trust formation. Most agents remain concentrated at the identity layer, rather than progressing toward the higher layers required for a functioning agent economy.

\begin{figure}[htbp]
    \centering

    \begin{subfigure}[t]{0.48\textwidth}
        \centering
        \includegraphics[width=\textwidth]{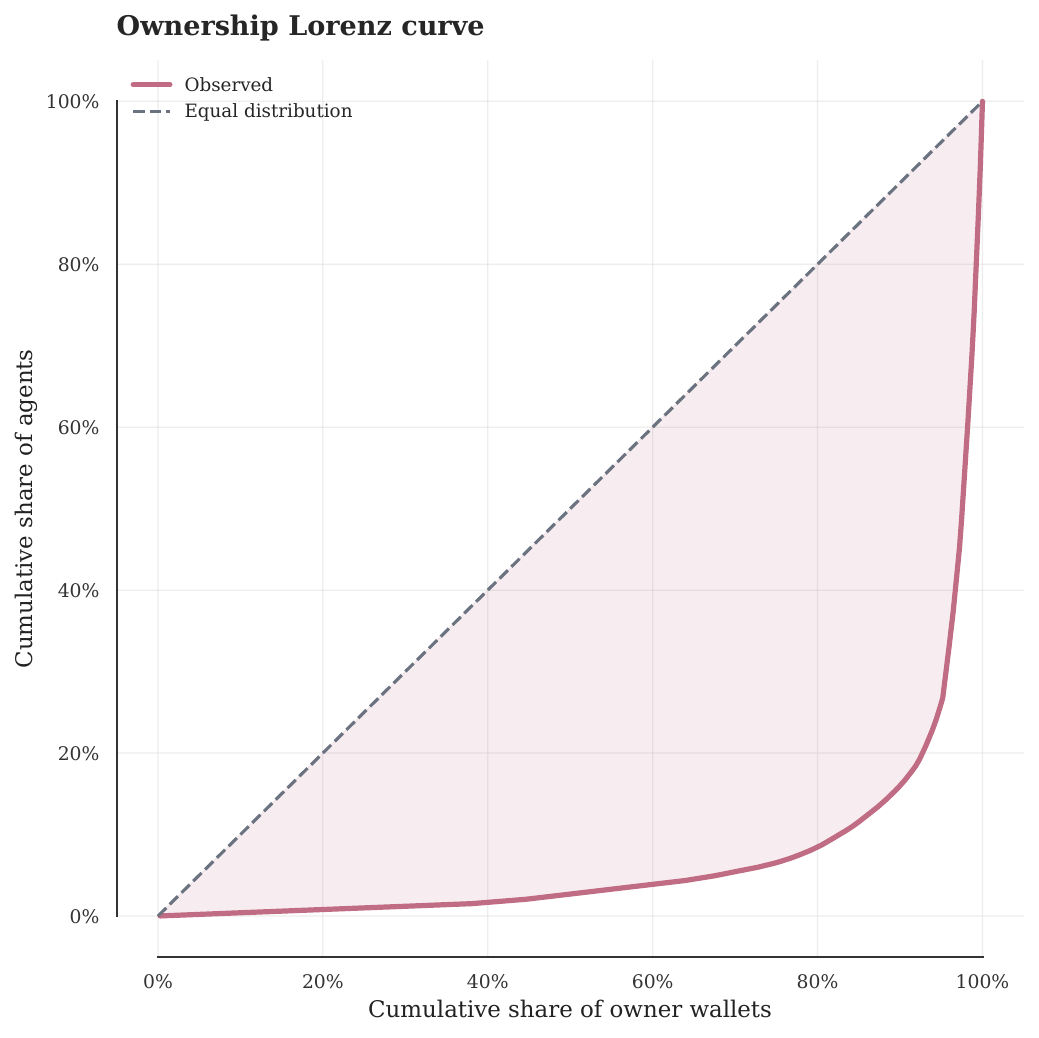}
        \caption{Agent ownership}
        \label{fig:owner_lorenz}
    \end{subfigure}
    \hfill
    \begin{subfigure}[t]{0.48\textwidth}
        \centering
        \includegraphics[width=\textwidth]{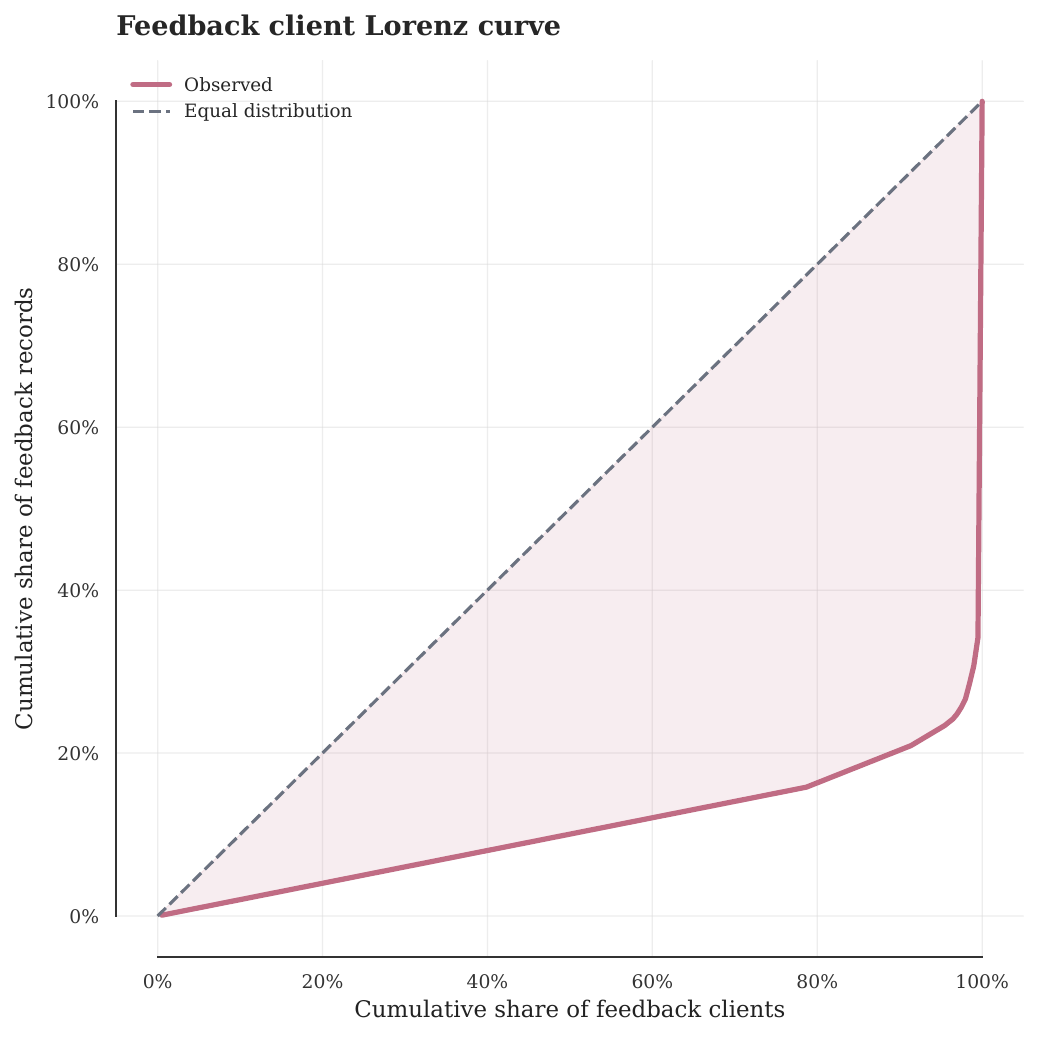}
        \caption{Feedback-client activity}
        \label{fig:client_lorenz}
    \end{subfigure}

    \caption{Lorenz curves for ERC-8004 ownership and reputation concentration. Panel (a) shows the concentration of agent ownership across owner wallets, while Panel (b) shows the concentration of reputation feedback across feedback-client addresses. Together, the figures indicate that both identity ownership and reputation activity are unevenly distributed in the early ERC-8004 ecosystem.}
    \label{fig:lorenz_curves}
\end{figure}

Figure~\ref{fig:readiness_funnel} presents the readiness funnel, which shows how the population of agents narrows as additional layers of evidence are required. The funnel structure provides a systematic way to evaluate whether ERC-8004 agents are moving beyond basic on-chain identity and toward operational participation. The results indicate that only a small subset of agents show evidence of richer metadata, service readiness, or feedback activity. This pattern suggests that many ERC-8004 registrations currently resemble inactive identities rather than economically active agents. The high level of ownership concentration reinforces this interpretation. Instead of being distributed across a broad base of end users, ERC-8004 agents appear to be held disproportionately by a small number of addresses, which may correspond to developers, early deployers, or speculative holders as shown in Figures~\ref{fig:owner_lorenz}.

% ============================================================
% Figure: Ownership and feedback-client Lorenz curves
% Draft labels:
% Main figure: fig:lorenz_curves
% Subfigures: fig:owner_lorenz and fig:client_lorenz
% Files:
% 05_ownership_lorenz_curve.pdf
% 13_feedback_client_lorenz_curve.pdf
% ============================================================

\begin{figure}[htbp]
    \centering
    \includegraphics[width=0.9\textwidth]{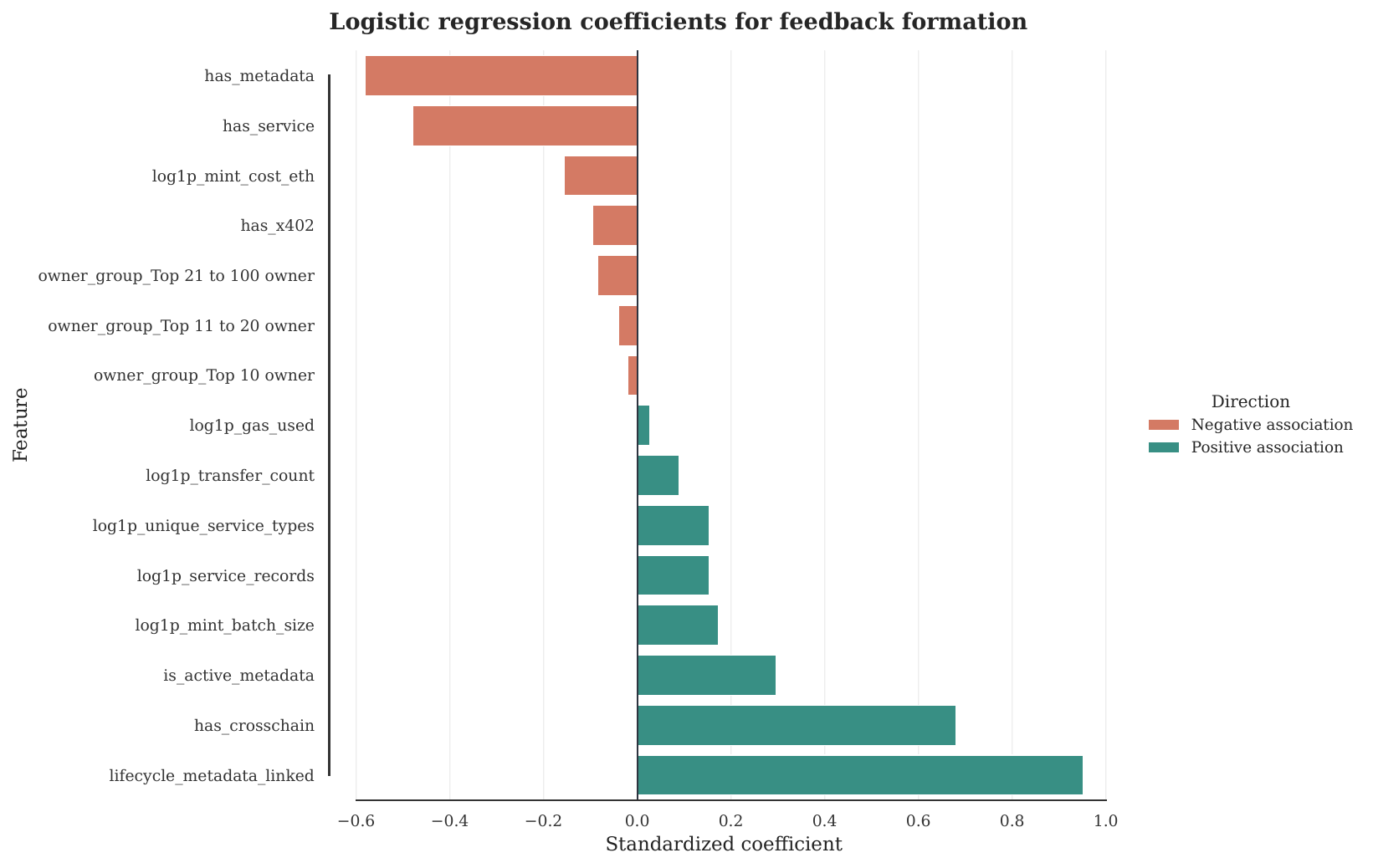}
    \caption{Standardized logistic regression coefficients associated with feedback formation. The model is exploratory and should be interpreted as evidence of association rather than causality.}
    \label{fig:regression_coefficients}
\end{figure}

Feedback activity provides additional evidence of limited ecosystem maturity. In a healthy agent economy, reputation and feedback mechanisms should support trust formation through broad, repeated, and diverse participation. In the current ERC-8004 ecosystem, however, feedback is sparse and highly concentrated, with a substantial share originating from a single address as shown in Figure~\ref{fig:client_lorenz}. Figure~\ref{fig:regression_coefficients} provides supplementary evidence on which observable agent characteristics are associated with entering the feedback layer. The strongest positive associations are observed for \texttt{lifecycle\_metadata\_linked} and \texttt{has\_crosschain}, suggesting that agents with broader observable infrastructure are more likely to receive feedback. Smaller positive associations are also observed for service intensity variables, including \texttt{log1p\_service\_records} and \texttt{log1p\_unique\_service\_types}. However, the binary indicators \texttt{has\_metadata} and \texttt{has\_service} have negative coefficients. This should not be interpreted as evidence that metadata or services reduce feedback. Rather, it likely reflects sparsity, overlap among readiness indicators, and the uneven structure of the early ERC-8004 dataset. Overall, the regression supports the descriptive interpretation that feedback formation is associated with richer observable infrastructure, but the relationship remains selective and should be interpreted as association rather than causality.

Overall, the descriptive evidence indicates that ERC-8004 is currently functioning more as an on-chain identity registry than as a mature infrastructure layer for an autonomous agent economy. At this stage, the ERC-8004 ecosystem appears to have established the lower layer of identity creation, but the higher layers of service activity, reputation formation, and economic coordination remain underdeveloped.

% ============================================================
% Figure: Logistic regression coefficient plot
% New label: fig:regression_coefficients
% File exists: 28_regression_coefficient_plot.pdf
% ============================================================

\subsection{Network Structure of ERC-8004 Agent Activity}

To complement the tabular readiness indicators, we construct four network views of the ERC-8004 agent ecosystem. These plots are designed to show whether agent activity is broadly distributed or concentrated around a small number of wallets, feedback clients, service domains, and transfer paths. The network analysis is exploratory and should be interpreted as structural evidence of ecosystem organization rather than proof of autonomous agent behavior.

\subsubsection{Owner-to-Agent Network}

\begin{figure}[htbp]
    \centering
    \includegraphics[width=0.95\textwidth]{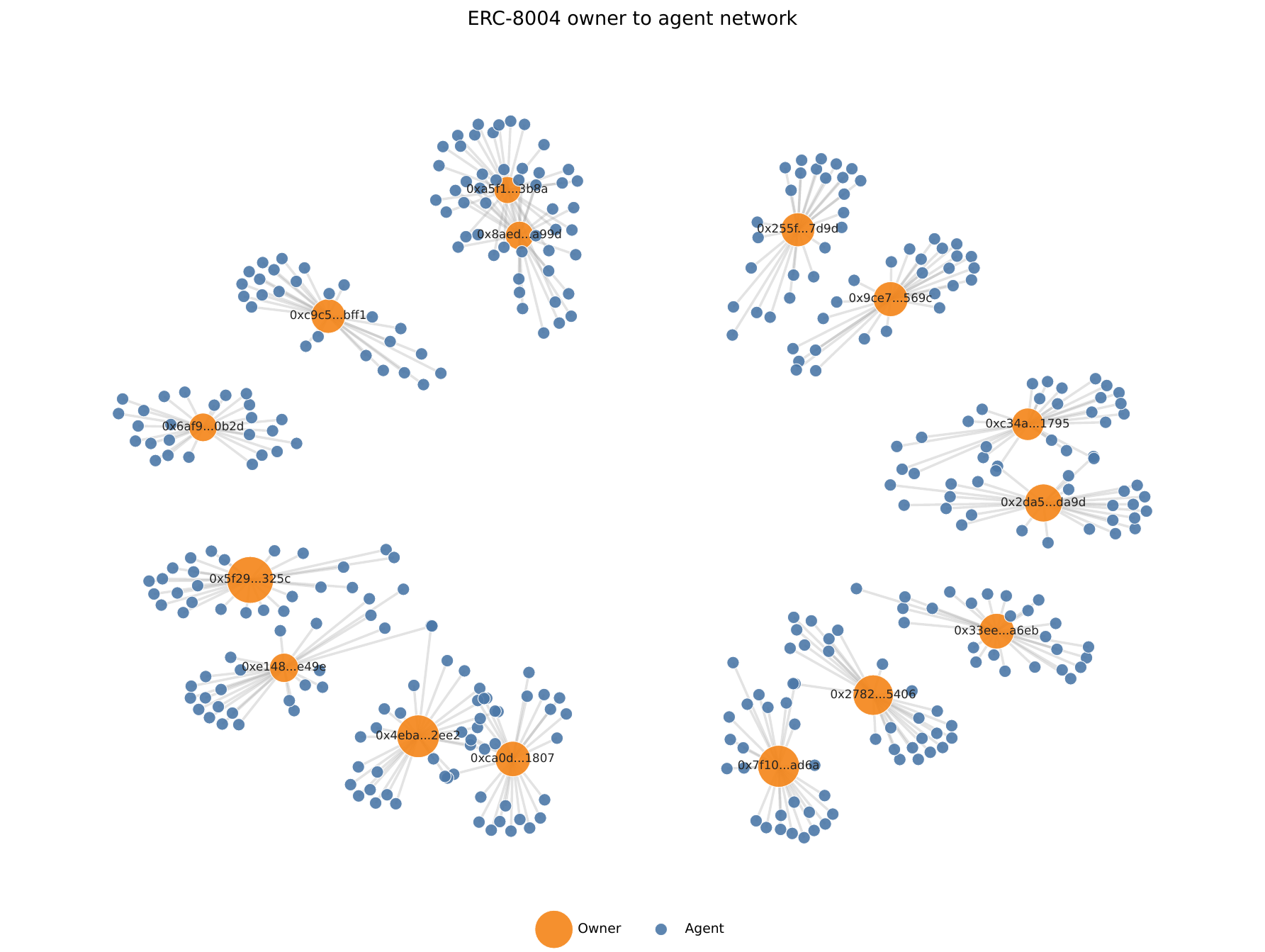}
    \caption{Owner-to-agent network for ERC-8004 agents. Orange nodes represent owner wallets and blue nodes represent agent identities. The visualization shows that many agent identities are connected to a small number of high-volume owner wallets, indicating concentrated ownership in the early ERC-8004 ecosystem.}
    \label{fig:owner_agent_network}
\end{figure}

Figure~\ref{fig:owner_agent_network} visualizes the relationship between owner wallets and ERC-8004 agent identities. Orange nodes represent owner wallets, while blue nodes represent agent IDs. The plot highlights a clear hub-and-spoke structure, where a small number of wallets are connected to many agents. This means that agent registration is not evenly distributed across the ecosystem. Instead, many agents appear to be held or deployed by a limited group of high-volume owner wallets. This pattern is important because a mature agent economy would ideally show broader participation across many independent owners, rather than relying heavily on a small number of concentrated deployers.

Table~\ref{tab:ownership_concentration} provides quantitative support for the network structure shown in Figure~\ref{fig:owner_agent_network}. The dataset contains 394 unique owner wallets for 10,000 registered agents, which already indicates that many agents are grouped under relatively few owners. The mean number of agents per owner is 25.38, but the median is only 3. This large gap between the mean and median shows that the ownership distribution is highly skewed: most wallets hold only a small number of agents, while a small number of wallets hold very large numbers of agents. The maximum value confirms this concentration, with the largest owner wallet holding 779 agents.
The concentration metrics further reinforce this interpretation. The ownership Gini coefficient is 0.863, indicating a highly unequal distribution of agent ownership across wallets. The HHI value of 0.034 also suggests non-trivial concentration, although ownership is not monopolized by a single wallet. Together, the network plot and the concentration table show that ERC-8004 identity formation is still dominated by a small set of high-volume owners or deployers. This does not necessarily imply harmful centralization, since early-stage ecosystems often involve developer wallets, testing wallets, or infrastructure providers. However, it does indicate that ERC-8004 adoption is not yet broadly distributed across a large base of independent participants.

% ============================================================
% Table: Ownership concentration
% Source CSV: 07_ownership_concentration_metrics.csv
% Draft label: tab:ownership_concentration
% ============================================================

\begin{table}[htbp]
\centering
\caption{Ownership concentration among ERC-8004 agent owner wallets.}
\label{tab:ownership_concentration}
\small
\begin{tabular}{lr}
\toprule
Metric & Value \\
\midrule
Unique owner wallets & 394 \\
Ownership HHI & 0.034283 \\
Ownership Gini & 0.863005 \\
Mean agents per owner & 25.380711 \\
Median agents per owner & 3 \\
Maximum agents owned by one wallet & 779 \\
\bottomrule
\end{tabular}
\end{table}

Overall, the owner-to-agent network shows that ERC-8004 identity registration is active but unevenly distributed. The presence of 10,000 registered agents may suggest rapid adoption at first glance, but the ownership structure reveals a more concentrated pattern. This finding supports the broader argument of the paper: early ERC-8004 activity is visible on-chain, but the ecosystem has not yet reached a mature stage of broad and distributed participation.

\subsubsection{Feedback Client-to-Agent Network}

\begin{figure}[htbp]
    \centering

    \begin{subfigure}[t]{0.48\textwidth}
        \centering
        \includegraphics[width=\textwidth]{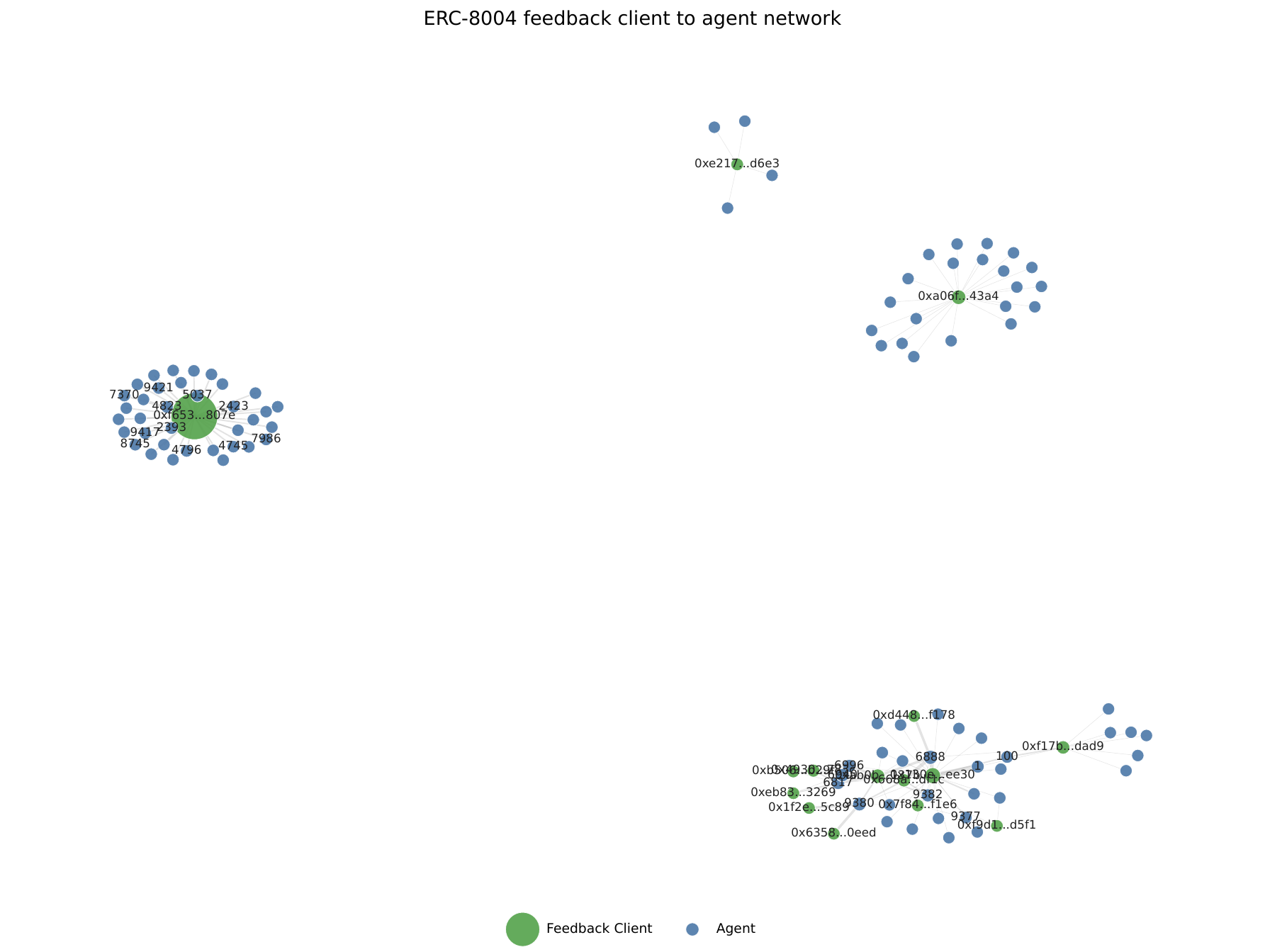}
        \caption{Feedback client-to-agent network}
        \label{fig:feedback_client_agent_network}
    \end{subfigure}
    \hfill
    \begin{subfigure}[t]{0.48\textwidth}
        \centering
        \includegraphics[width=\textwidth]{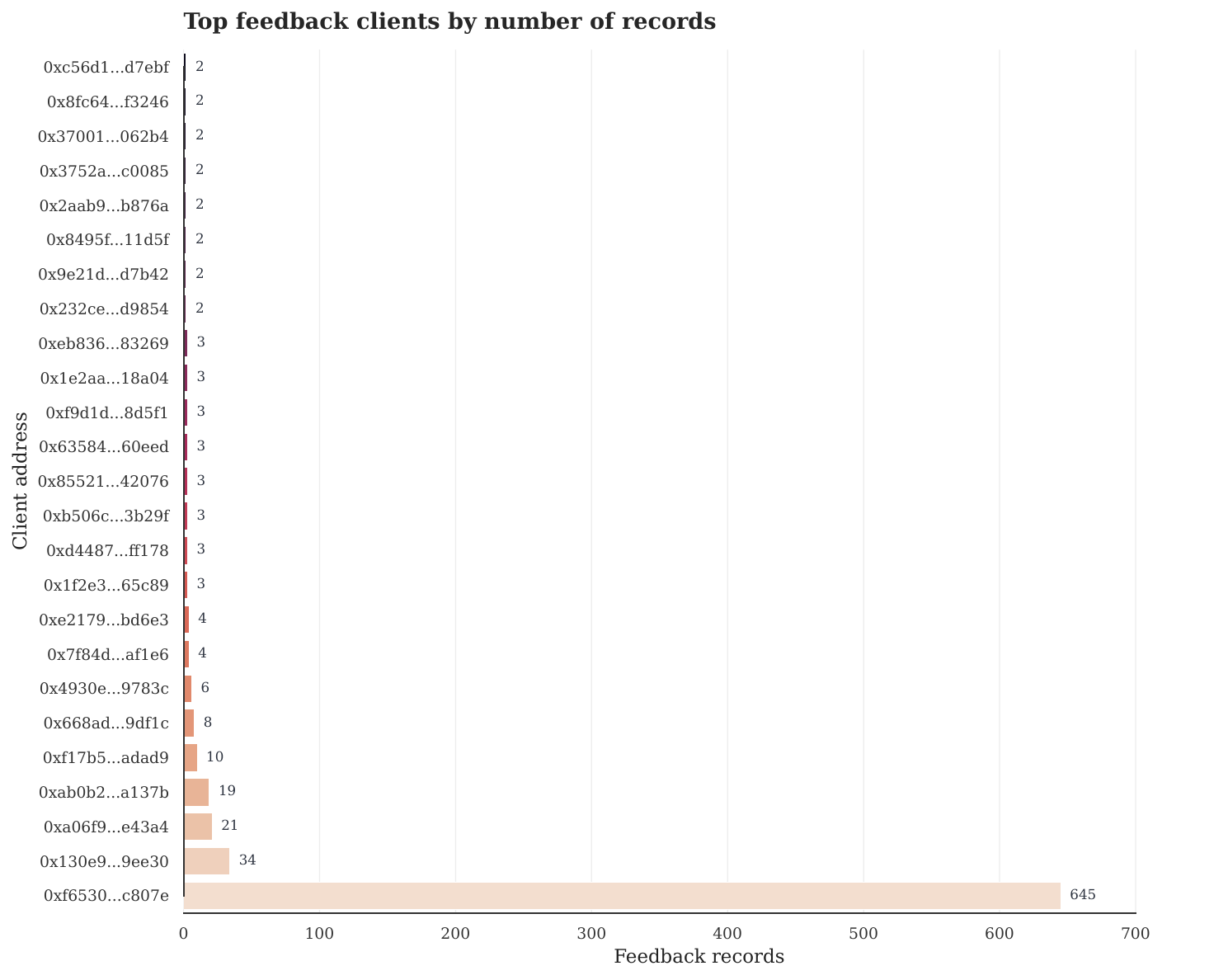}
        \caption{Top feedback clients}
        \label{fig:feedback_clients}
    \end{subfigure}

    \caption{Feedback concentration in the ERC-8004 reputation layer. Panel (a) shows the network between feedback clients and agents, where green nodes represent feedback client addresses and blue nodes represent agents. Panel (b) reports the most active feedback clients. Together, the figures show that reputation activity is concentrated around a small number of client addresses, suggesting that feedback formation remains limited and uneven in the early ERC-8004 ecosystem.}
    \label{fig:feedback_network_and_clients}
\end{figure}

Figure~\ref{fig:feedback_network_and_clients} examines the structure of reputation activity in the ERC-8004 ecosystem. Panel~(a) visualizes the relationship between feedback clients and agents that received reputation feedback. Green nodes represent feedback client addresses, while blue nodes represent agent identities. The network shows a highly uneven structure: a small number of feedback clients are connected to many agents, while most agents have limited or no visible feedback relationships. This indicates that reputation formation is not yet broadly distributed across many independent reviewers.

Panel~(b) complements the network visualization by showing the most active feedback clients. The concentration is substantial. Feedback records are observed for only 628 agents, equivalent to 6.28\% of the 10,000 registered agents. Moreover, the majority of feedback activity originates from a single client address, which contributes 645 records, or 65.8\% of all feedback entries. The top five feedback clients together account for 92.4\% of feedback records. This pattern suggests that the reputation layer is currently driven by a very small number of high-activity clients rather than by broad ecosystem participation.

% ============================================================
% Table: Feedback summary
% Source CSV: 14_feedback_summary.csv
% Draft label: tab:feedback_summary
% ============================================================

\begin{table}[htbp]
\centering
\caption{Summary of ERC-8004 reputation feedback records.}
\label{tab:feedback_summary}
\small
\begin{tabular}{lr}
\toprule
Metric & Value \\
\midrule
Total feedback records & 980 \\
Agents with feedback records & 628 \\
Unique feedback clients & 197 \\
Largest client feedback records & 645 \\
Largest client share (\%) & 65.82 \\
Feedback client HHI & 0.435800 \\
Feedback client Gini & 0.782513 \\
Agents with more than one feedback client & 18 \\
\bottomrule
\end{tabular}
\end{table}

This concentration is also reflected in the feedback-provider concentration metrics. The Herfindahl-Hirschman Index for feedback clients is 0.436, while the Gini coefficient is 0.783. These values indicate a highly unequal distribution of feedback activity across client addresses. In practical terms, this means that although the ERC-8004 Reputation Registry is being used, its early use remains narrow and centralized around a few dominant feedback providers. From an operational readiness perspective, this is important because reputation systems become more meaningful when feedback comes from diverse, repeated, and independent counterparties. A reputation layer dominated by a small number of clients may reflect early testing, developer-driven interaction, or platform-specific feedback processes rather than mature market-wide trust formation.

% ============================================================
% Figure: Feedback network and top feedback clients
% Main label: fig:feedback_network_and_clients
% Subfigure labels:
% fig:feedback_client_agent_network
% fig:feedback_clients
% ============================================================

Overall, the feedback network reinforces the broader finding that ERC-8004 has not yet developed a mature and distributed reputation layer. The existence of feedback records indicates that the Reputation Registry is operational at a basic level, but the limited agent coverage and extreme client concentration suggest that reputation formation remains early-stage. This finding is consistent with the paper's broader argument that ERC-8004 currently functions more strongly as an identity and registration infrastructure than as a fully developed agent economy with distributed trust formation.

\subsubsection{Wallet Transfer Network}

\begin{figure}[htbp]
    \centering
    \includegraphics[width=0.95\textwidth]{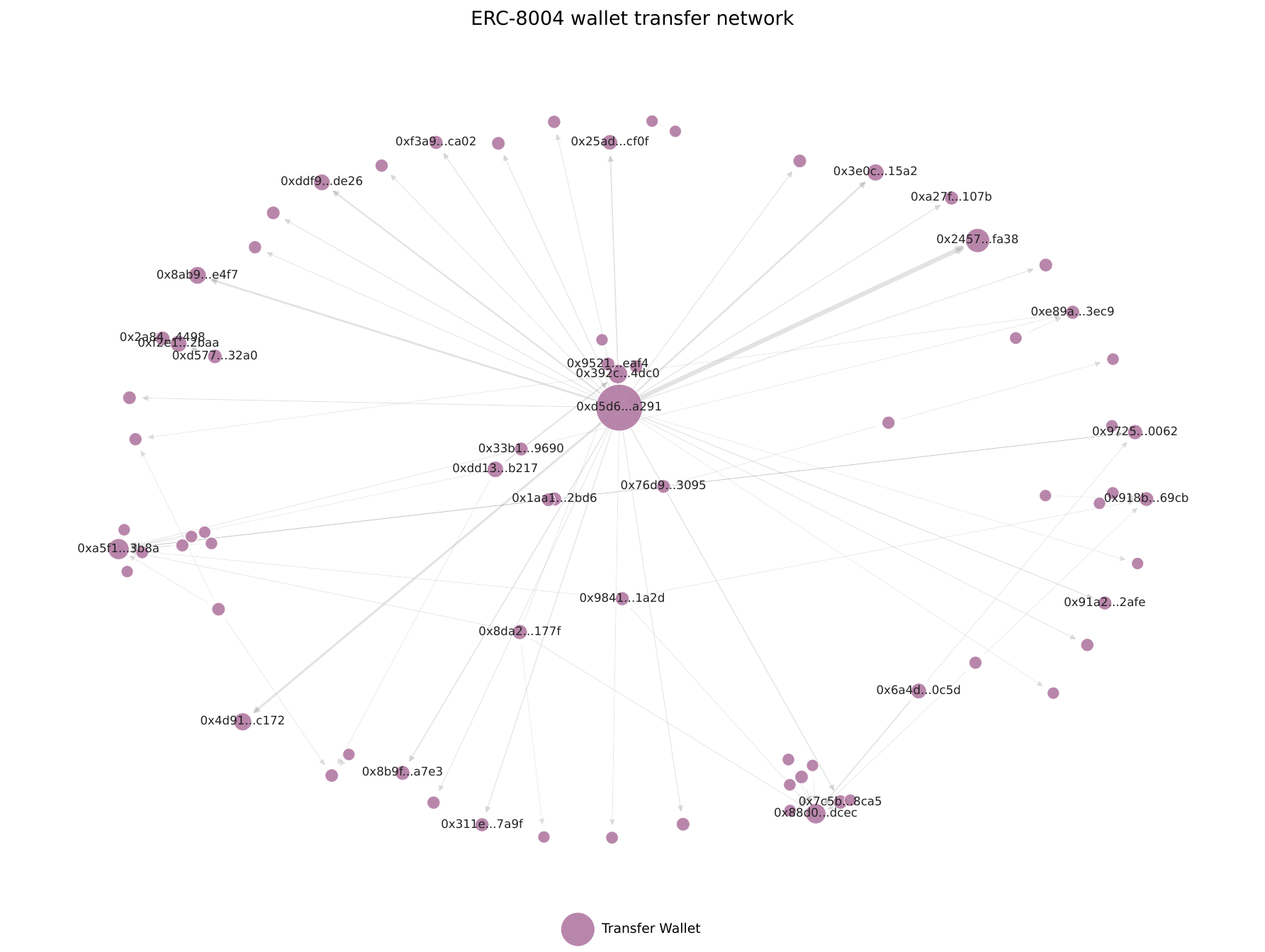}
    \caption{Wallet transfer network for ERC-8004 agent identities. Purple nodes represent wallets and directed edges represent observed transfer relationships. The figure captures ownership movement across wallets, but transfer activity should not be interpreted as proof of autonomous agent behavior.}
    \label{fig:wallet_transfer_network}
\end{figure}

Figure~\ref{fig:wallet_transfer_network} visualizes wallet-to-wallet transfer relationships observed in the transfer history table. Purple nodes represent wallets and directed edges represent observed transfers between wallets. The graph provides a structural view of agent ownership movement, but it should not be interpreted as evidence of agent autonomy. Transfers may be initiated by human operators, scripts, marketplaces, or infrastructure processes. The main insight is that transfer activity is observable but limited in interpretability. It provides evidence of movement in ownership or custody, but not direct evidence that agents are performing independent economic actions.

\subsubsection{Combined Agent Evidence Network}

\begin{figure}[htbp]
    \centering
    \includegraphics[width=0.98\textwidth]{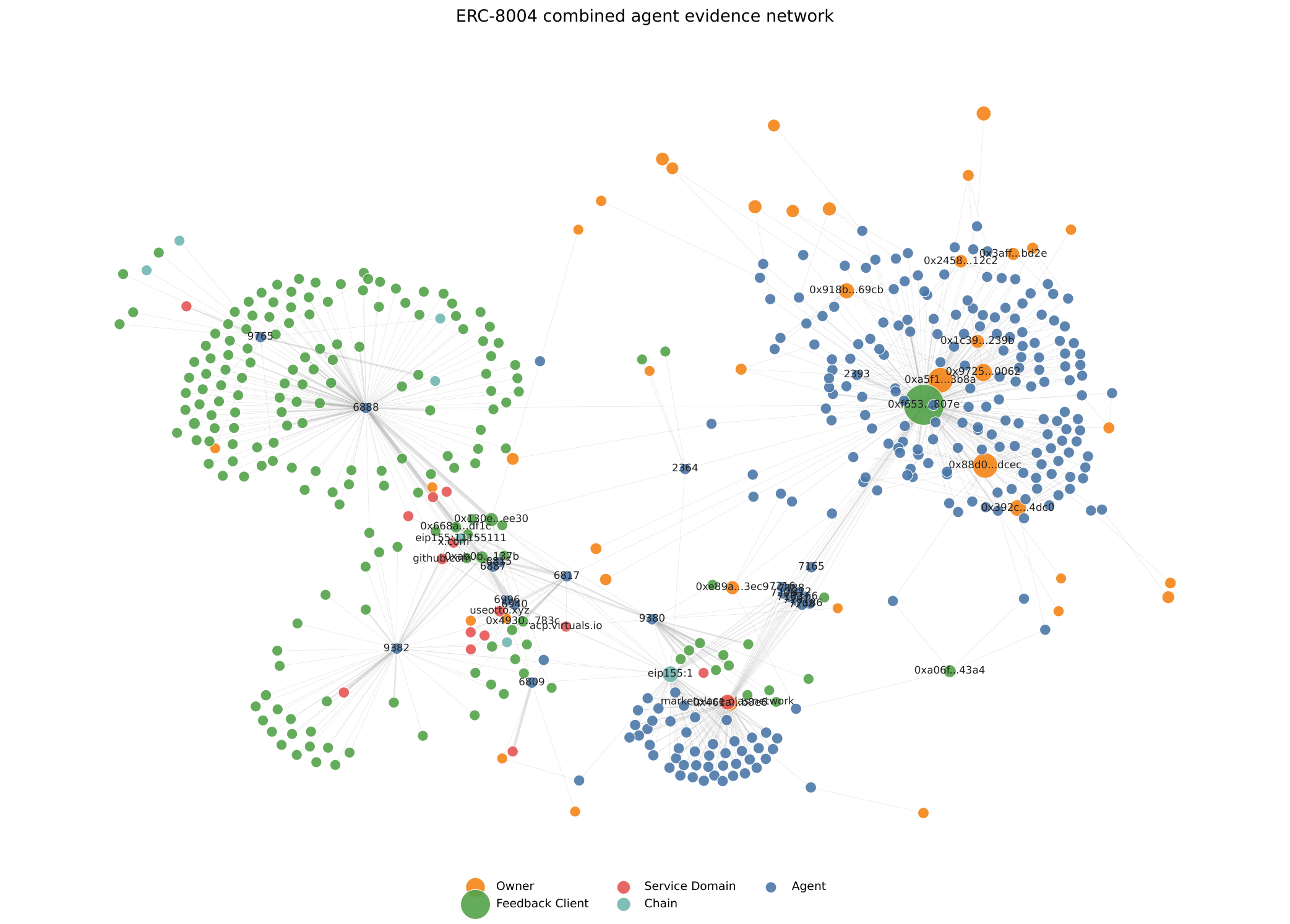}
    \caption{Combined agent evidence network for ERC-8004 agents. The network links agents to owner wallets, feedback clients, service endpoint domains, and cross-chain namespaces. The figure shows that richer operational evidence is concentrated among a small subset of agents.}
    \label{fig:combined_agent_evidence_network}
\end{figure}

Figure~\ref{fig:combined_agent_evidence_network} presents the combined agent evidence network, which integrates several observable readiness layers into a single structural view. In this network, agent nodes are connected to owner wallets, feedback clients, service endpoint domains, and cross-chain namespaces whenever these relationships are observed in the dataset. This visualization is useful because it moves beyond counting agents in separate readiness categories and instead shows how different evidence layers cluster around specific agents.

The network reveals that richer operational evidence is concentrated among a small subset of agents. Only a limited number of agents are connected simultaneously to multiple evidence sources, such as ownership, feedback, service domains, and cross-chain registration. Most registered agents do not appear prominently in this richer network because they lack one or more of the observable layers required for multi-dimensional participation. This pattern is consistent with the readiness funnel and supports the main empirical finding that ERC-8004 adoption is registration-heavy but operationally shallow. In other words, while the registry contains many agent identities, only a small group of agents shows evidence of broader ecosystem engagement.

\begin{figure}[htbp]
    \centering

    \begin{subfigure}[t]{0.48\textwidth}
        \centering
        \includegraphics[width=\textwidth]{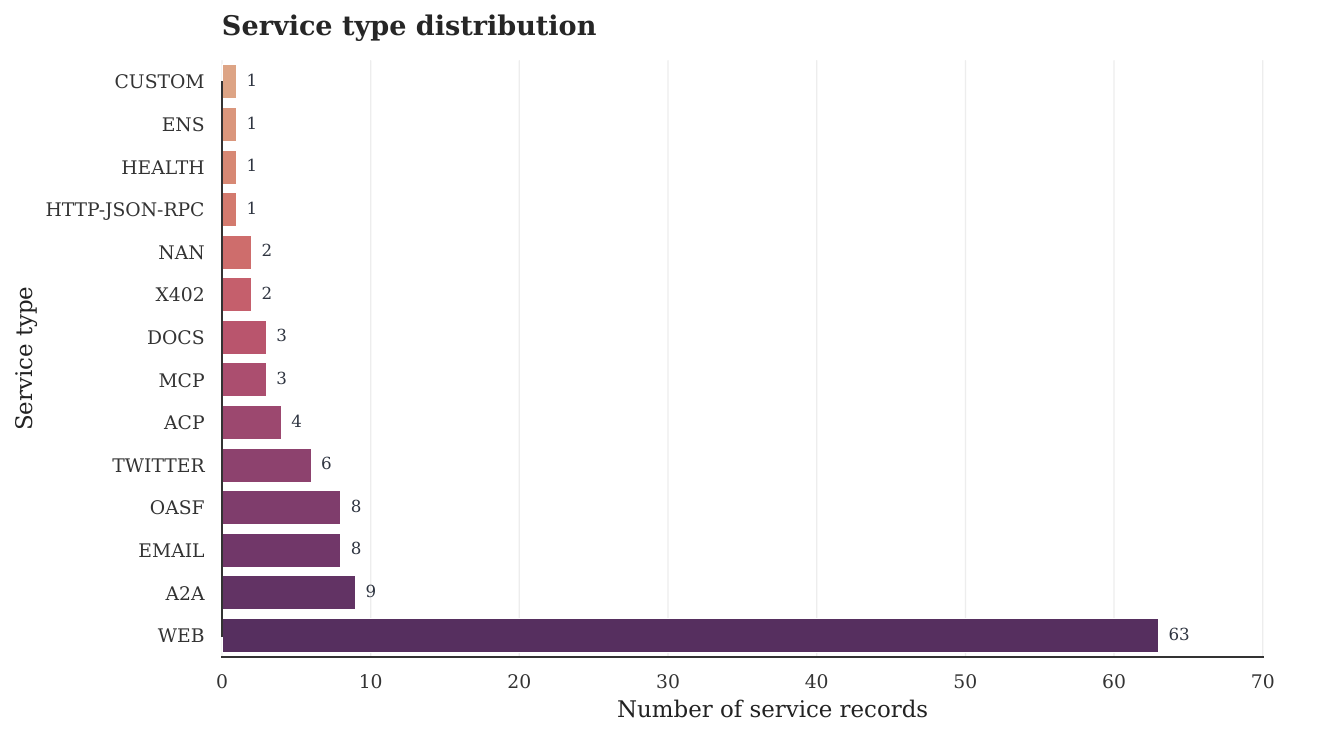}
        \caption{Service type distribution}
        \label{fig:service_types}
    \end{subfigure}
    \hfill
    \begin{subfigure}[t]{0.48\textwidth}
        \centering
        \includegraphics[width=\textwidth]{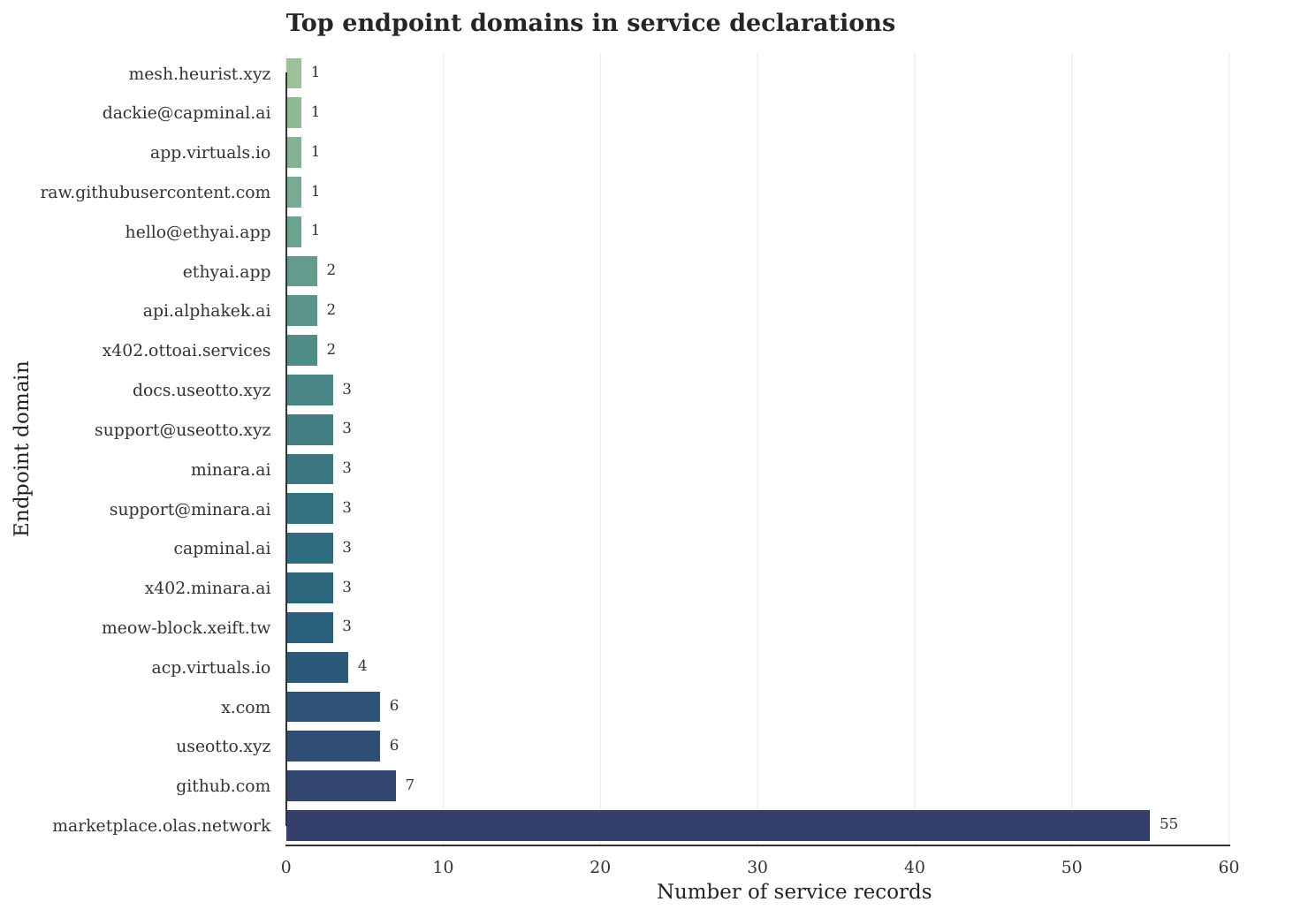}
        \caption{Top endpoint domains}
        \label{fig:endpoint_domains}
    \end{subfigure}

    \caption{Declared service infrastructure among ERC-8004 agents with service records. Panel (a) reports the distribution of service types, showing that web endpoints dominate the observed service layer. Panel (b) reports the most common endpoint domains, showing that service declarations are concentrated around a small number of domains. Together, the figures indicate that the service layer is present but remains limited in diversity and depth.}
    \label{fig:service_and_domains}
\end{figure}

Figure~\ref{fig:service_and_domains} summarizes both the distribution of declared service types and the endpoint domains used in service metadata. Panel~(a) shows that web endpoints dominate the service layer, accounting for 56.3\% of service records. Other service categories appear much less frequently: A2A accounts for 8.0\%, EMAIL for 7.1\%, OASF for 7.1\%, MCP for 2.7\%, X402 for 1.8\%, and HTTP-JSON-RPC for only 0.9\%. Since MCP refers to the Model Context Protocol and A2A refers to agent-to-agent interaction, their low representation suggests that most service declarations do not yet reflect a mature inter-agent communication layer.

Panel~(b) shows the endpoint-domain distribution. The most common domain is \\ \texttt{marketplace.olas.network}, followed by smaller clusters of domains such as \texttt{github.com}, \texttt{x.com}, \texttt{useotto.xyz}, \texttt{acp.virtuals.io}, and other project-specific endpoints. This distribution suggests that service metadata is not yet widely diversified across many independent infrastructure providers. Instead, a small number of domains account for a large share of declared endpoints. The service layer therefore appears to be present but narrow, with many service records pointing to web or platform-specific endpoints rather than to a broad ecosystem of operational agent interfaces.

% ============================================================
% Figure: Service type distribution and endpoint domains
% Main label: fig:service_and_domains
% Subfigure labels:
% fig:service_types
% fig:endpoint_domains
% ============================================================

Taken together, the combined evidence network and service-layer figures provide a structural view of ERC-8004's current maturity. The combined network shows that only a small subset of agents connects across multiple observable readiness layers. The service-type and endpoint-domain distributions show that even among agents with service records, declared services are concentrated in a small number of formats and domains. These findings reinforce the interpretation that ERC-8004 currently functions more strongly as an identity-registration layer than as a mature agent economy with broad service discoverability, distributed trust formation, and multi-party interaction.

Overall, the network analysis complements the readiness funnel by showing not only how many agents reach each layer, but also how those layers are connected. The owner-to-agent network indicates concentrated identity ownership, the feedback client-to-agent network indicates concentrated reputation activity, and the combined evidence network shows that multi-layer readiness is limited to a small portion of the ecosystem. This does not mean that ERC-8004 lacks potential. Rather, it shows that the current empirical footprint of ERC-8004 remains early-stage: identity registration is visible at scale, but the service, reputation, and interaction layers needed for a mature agent economy remain sparse and unevenly distributed.

\section{Limitations}

This study has several limitations. First, the analysis is based on an early snapshot of ERC-8004 deployment on Ethereum mainnet, covering agent registrations from January to April 2026. Since ERC-8004 is still an emerging standard, the observed patterns may change as the specification matures, tooling improves, and additional agents are registered. The results should therefore be interpreted as a baseline assessment of early adoption rather than a definitive evaluation of the long-term ERC-8004 ecosystem.

Second, the analysis is limited to observable on-chain records and metadata that could be retrieved and parsed from the available dataset. These records allow us to measure registration, metadata availability, service declarations, feedback, transfers, cross-chain registration, and network structure. Although the dataset contains 10,000 registered agents, only a small subset has fully parsed metadata and declared services. As a result, inferences about agent capabilities, service readiness, and endpoint diversity are necessarily limited. Similarly, feedback records are available for only a minority of agents and are highly concentrated among a small number of client addresses. This limits the extent to which current feedback activity can be interpreted as broad reputation formation across the ecosystem.

Finally, the regression and network analyses are descriptive and exploratory. The logistic regression identifies agent characteristics associated with receiving feedback, but it does not establish causal relationships. For example, metadata linkage may make an agent more likely to receive feedback, but it is also possible that more visible or actively promoted agents are both more likely to have metadata and more likely to receive feedback. Similarly, the network analysis reveals concentration and clustering, but it cannot determine the off-chain processes that generate those structures. Despite these limitations, the study provides a useful empirical baseline for monitoring how ERC-8004 evolves from identity registration toward richer forms of operational agent infrastructure.

\section{Conclusion}

This paper examined whether ERC-8004 agents on Ethereum demonstrate operational readiness beyond basic identity registration. Using a dataset of the first 10,000 ERC-8004 agents, we constructed an agent-level feature table covering identity status, metadata, service declarations, reputation feedback, transfers, cross-chain registration, mint economics, and network relationships. We then proposed a layered readiness framework to assess whether registered agents move from on-chain identity toward discoverability, service exposure, reputation formation, and broader ecosystem participation.

The findings show that early ERC-8004 adoption is registration-heavy but operationally shallow. While the identity layer has reached visible scale, higher readiness layers remain limited. Only a small share of agents have parsed metadata, service records, feedback, or cross-chain registrations, and only 19 from 10,000 agents combine metadata, services, feedback, and cross-chain registration. Ownership and reputation activity are also highly concentrated, with a small number of wallets and feedback clients accounting for a large share of observed activity. The network analysis reinforces this conclusion by showing that richer operational evidence clusters around a small subset of agents rather than being broadly distributed across the ecosystem.

These results suggest that ERC-8004 currently functions more strongly as an identity-registration layer than as a mature agent economy. This does not diminish the importance of the standard. On the contrary, ERC-8004 provides a useful foundation for representing agents, linking them to metadata, and building reputation. However, the empirical evidence indicates that the transition from agent identity to agent economy remains incomplete. A functioning agent economy will require more than registered identities. It will require reliable metadata, active service endpoints, diverse reputation formation, and distributed relationships among agents, users, wallets, and service providers.

The paper contributes to the literature on decentralized AI agents and blockchain-based agent economies by offering an empirical framework for measuring operational readiness. Rather than assuming that agent registration implies agent functionality, the framework distinguishes between identity creation and observable evidence of use. This distinction is important for researchers, developers, and standard designers because it highlights where current infrastructure is already visible and where substantial gaps remain. As ERC-8004 and related standards evolve, ongoing empirical measurement will be essential for tracking whether blockchain-registered agents become active participants in decentralized economies or remain primarily static identity tokens.

\newpage
\bibliographystyle{abbrvnat}
\bibliography{references}

\end{document}